\documentclass{aastex63}

\usepackage{CJK}
\usepackage{booktabs}
\usepackage{multirow}
\usepackage{amsmath}
\shorttitle{Laser propagation noise for TQ}
\shortauthors{Su et al.}
\begin{document}
\begin{CJK*}{UTF8}{gbsn}
	
\title{Analyses of Laser Propagation Noises for TianQin Gravitational Wave Observatory Based on the Global Magnetosphere MHD Simulations}

\correspondingauthor{Wang Y., Zhou C., Lu L.}
\email{ywang12@hust.edu.cn, chenzhou@whu.edu.cn, lulingfeng@swip.ac.cn}

\author{Wei Su (苏威)}
\affiliation{MOE Key Laboratory of Fundamental Physical Quantities Measurements, Hubei Key Laboratory of Gravitation and Quantum Physics, PGMF, Department of Astronomy, and School of Physics, Huazhong University of Science and Technology, Wuhan 430074, China}

\author{Yan Wang (王炎)}
\affiliation{MOE Key Laboratory of Fundamental Physical Quantities Measurements, Hubei Key Laboratory of Gravitation and Quantum Physics, PGMF, Department of Astronomy, and School of Physics, Huazhong University of Science and Technology, Wuhan 430074, China}

\author{Chen Zhou (周晨)}
\affiliation{Department of Space Physics, School of Electronic Information, Wuhan University, Wuhan 430072, China}

\author{Lingfeng Lu}
\affiliation{Southwestern Institute of Physics, Chengdu 610041, China} 

\author{Ze-Bing Zhou (周泽兵)}
\affiliation{MOE Key Laboratory of Fundamental Physical Quantities Measurements, Hubei Key Laboratory of Gravitation and Quantum Physics, PGMF, Department of Astronomy, and School of Physics, Huazhong University of Science and Technology, Wuhan 430074, China}

\author{T.M. Li (李汤姆)}
\affiliation{School of Astronomy and Space Science, Nanjing University, Nanjing 210023, China}
\affiliation{Key Laboratory of Modern Astronomy \& Astrophysics, Nanjing University, China}

\author{Tong Shi (施通)}
\affiliation{Department of Climate and Space Sciences and Engineering, University of Michigan, Ann Arbor, MI 48109, USA}

\author{Xin-Chun Hu (胡新春)}
\affiliation{MOE Key Laboratory of Fundamental Physical Quantities Measurements, Hubei Key Laboratory of Gravitation and Quantum Physics, PGMF, Department of Astronomy, and School of Physics, Huazhong University of Science and Technology, Wuhan 430074, China}

\author{Ming-Yue Zhou (周明越)}
\affiliation{MOE Key Laboratory of Fundamental Physical Quantities Measurements, Hubei Key Laboratory of Gravitation and Quantum Physics, PGMF, Department of Astronomy, and School of Physics, Huazhong University of Science and Technology, Wuhan 430074, China}

\author{Ming Wang (王明)}
\affiliation{Institute of Space Weather, School of Mathematics and Statistics, Nanjing University of Information Science and Technology, Nanjing 210044, China}

\author{Hsien-Chi Yeh}
\affiliation{TianQin Research Center for Gravitational Physics \& School of Physics and Astronomy, 
	Sun Yat-sen University, Zhuhai 519082, China} 

\author{Han Wang}
\affiliation{MOE Key Laboratory of Fundamental Physical Quantities Measurements, Hubei Key Laboratory of Gravitation and Quantum Physics, PGMF, Department of Astronomy, and School of Physics, Huazhong University of Science and Technology, Wuhan 430074, China}
\affiliation{TianQin Research Center for Gravitational Physics \& School of Physics and Astronomy, 
	Sun Yat-sen University, Zhuhai 519082, China} 

\author{P.F. Chen (陈鹏飞)}
\affiliation{School of Astronomy and Space Science, Nanjing University, Nanjing 210023, China}
\affiliation{Key Laboratory of Modern Astronomy \& Astrophysics, Nanjing University, China}

%\collaboration{1}{(AAS Journals Data Scientists collaboration)}

%\nocollaboration{1}

%\altaffiliation{AASTeX v6+ programmer}

%\collaboration{1}{(LaTeX collaboration)}




\begin{abstract}

TianQin is a proposed space-borne gravitational wave (GW) observatory composed of three identical satellites orbiting around the geocenter with a radius of $10^5$ km. It aims at detecting GWs in the frequency range of 0.1 mHz -- 1 Hz. The detection of GW relies on the high precision measurement of optical path length at $10^{-12}$~m level. 
The dispersion of space plasma can lead to the optical path difference (OPD, $\Delta l$) along the propagation of laser beams between any pair of satellites.
Here, we study the OPD noises for TianQin.
The Space Weather Modeling Framework is used to simulate the interaction between the Earth magnetosphere and solar wind. From the simulations, we extract the magnetic field and plasma parameters on the orbits of TianQin at 
four relative positions of the satellite constellation in the Earth magnetosphere. 
We calculate the OPD noise for single link, Michelson combination, and Time-Delay Interferometry (TDI) combinations ($\alpha$ and $X$). For single link and Michelson interferometer, the maxima of $|\Delta l|$ are on the order of 1 pm. For the TDI combinations, 
these can be suppressed to about 0.004 and 0.008 pm for $\alpha$ and $X$.
The OPD noise of the Michelson combination is colored in the concerned frequency range; 
while the ones for the TDI combinations are approximately white. 
Furthermore, we calculate the ratio of the equivalent strain of the OPD noise to that of TQ, 
and find that the OPD noises for the TDI combinations 
can be neglected in the most sensitive frequency range of TQ.

\end{abstract}

\keywords{plasmas, gravitational waves, Sun: solar wind}


\section{Introduction} \label{sec:intro}

%Since the direct detection of the gravitational waves (GWs) from the merger of a pair of stellar mass 
%binary black holes (GW150914) by the two advanced detectors 
%of the Laser Interferometer Gravitational-Wave Observatory (LIGO) \cite{LIGO2016}, 
%there are more than ten GW events have been detected by advanced LIGO and advanced 
%Virgo \cite{2016PhRvL.116x1103A, 2017PhRvL.119n1101A, 2017PhRvL.118v1101A, 
%	2017ApJ...851L..35A, 2017PhRvL.119p1101A, 2019PhRvX...9c1040A}. 
%Recently, KAGRA \cite{Somiya_2012} has joined in the ground-based GW detector network. 
%Due to the unshieldable impacts from the environment, namely seismic noise and gravity gradient noise, 
%it is very difficult for these terrestrial laser interferometers to detect GWs with the frequencies lower than 10 Hz. 
%However, in the low frequencies, there are rich sources of GWs that can be used to study 
%the fundamental physics, astrophysics and cosmology \cite{2009LRR....12....2S}. 
%The aim of space-borne laser interferometers is to explore the GWs in millihertz range (0.1 mHz--1 Hz). 
%Several projects, e.g. LISA \cite{LISA2017}, TianQin (TQ) \cite{Luo2016}, 
%DECIGO \cite{2011CQGra..28i4011K}, ASTROD-GW \cite{1998grco.conf..309N}, g-LISA \cite{Tinto2015gLISA}, 
%Taiji (ALIA descoped) \cite{TaiJi2015} and BBO \cite{Cutler2006BBO} have been proposed 
%and are currently under different stages of study and development. 

The first direct detection of gravitational waves (GWs) by the advanced Laser Interferometer Gravitational-Wave Observatory (LIGO)  opens up the era of the GW astronomy \citep{LIGO2016}. 
So far, more than fifty GW events generated by the coalescences of stellar-mass black hole binaries and double neutron stars have been detected by the advanced LIGO and advance Virgo \citep{2019PhRvX...9c1040A,2020arXiv201014527A}. 
The underground and cryogenic detector KAGRA \citep{Somiya_2012} has recently started joint observations
with the advanced LIGO and advance Virgo. 
Due to the terrestrial noises,  the ground-based detectors are most sensitive to the GW 
signals in the acoustic band ($\gtrsim$ 10 Hz). 

Several space-borne missions, e.g., LISA \citep{LISA2017},  TianQin \citep[{TQ};][]{Luo2016}, 
Taiji \citep[{ALIA descoped};][]{TaiJi2015}, ASTROD-GW \citep{1998grco.conf..309N}, gLISA \citep{Tinto2015gLISA},
BBO \citep{Cutler2006BBO} and DECIGO \citep{DECIGO2011}, have been proposed to explore the abundant GWs sources in the mHz band, which can be used to deepen our understandings in fundamental physics, astrophysics and cosmology.

Both LISA and TQ are in nearly equilateral triangular constellations which are formed by three drag-free satellites  
interconnected by infrared laser beams. 
The heterodyne transponder-type laser interferometers are used to measure the relative 
displacements of the test masses (TMs) with the accuracy of $10^{-12}~\rm{m/Hz^{1/2}}$ in mHz. 
This constellation forms up to three Michelson-type interferometers.
%These three satellites form a laser interference link, and the laser interference method is used to detect the change in distance caused by gravitational waves between the inspection masses on the satellites. 
Different from LISA, TQ's satellites will be deployed in a geocentric orbit with an 
altitude of $10^5$ km from the geocenter and the distances between each pair of 
satellites $\approx$ $1.7\times10^5$ km \citep{Luo2016}. 
The detector's plane formed by three satellites is optimized to detect the continuous GW signals from 
the candidate ultracompact white-dwarf binary RX J0806.3+1527 \citep{Israel2002}. 
Currently, both science cases \citep{Feng2019, 2019PhRvD.100d3003W, 2019PhRvD.100d4036S, 2019PhRvD.100h4024B, PhysRevD.101.103027, PhysRevD.102.063016, PhysRevD.102.063021} and technological realizations \citep{2020CQGra..37r5013L, 2019IJMPD..2850121Y, 2020IJMPD..2950056T, 2020CQGra..37k5005Y,Su2020A,2020PlST...22k5301L} have been under intensive investigations for TQ. 
A brief summary of TQ's recent progress can be found in \citet{Mei2020ptep}. 

%In space, the terrestrial noises (e.g., seismic noise and gravity gradient noise) 
%are much lower than those on the ground. 
%Instead, the space plasma will serve as the main source of environmental noises.
%One example is the dispersion effect induced by the space plasma 
%when the laser beams propagate from one spacecraft to the other \cite{Hutchinson2005}. 
%It can cause time-varying optical paths and time delays, hence affects the displacement 
%measurement accuracy. 
%Taking the typical magnitudes of electron number density and magnetic field in the Earth magnetosphere 
%as 1--10 cm$^{-3}$ and 1--10 nT, respectively, at the altitude of TQ's spacecraft, 
%one can see that the dominating factor of the dispersion is from the former one. 

The space plasma contributes as one of the main sources of environmental noises for space-borne GW detectors. 
For example, when the laser beams propagate in the space plasma 
the dispersion effect can lead to time delay and optical path difference (OPD) between different beams 
and produce additional noise for the relative displacement measurement. 
Since the plasma frequency $\omega_p$ in the space environment 
is much larger than the electron gyrofrequency, OPD is mainly caused by the total electron content (TEC) along each laser beam \citep{2020JA028579}. 
Moreover, the space magnetic field can induce the time variation of the polarization of electromagnetic (EM) waves, and the interaction between the space magnetic field and the test masses can generate additional non-conservative forces on the test masses  \citep{Hanson2003,Schumaker2003,Su2020A, LPF-M-mnras2020}. 

Space environment parameters, e.g., the magnetic field and density, vary significantly in time and space \citep{Su2016,Wang2018}, which can be categorized, in descending characteristic sizes, into three spatial scales: global scale, magnetohydrodynamic (MHD) scale, and plasma scale. 
In the global scale, the solar wind interacts with the Earth's magnetic dipole field to form structures such as bow shocks, magnetoheath, magnetopause, magnetotail, etc. \citep{Lv2015,Wang2016}. The density and magnetic field in different structures are essentially different. 
As the satellites orbiting around the Earth, the laser beam between two satellites passes through different structures, thus the OPD caused by the space plasma will be a function of time. Furthermore, because the solar wind is changing continually, the shapes and properties (e.g., the magnetic field and density) of these structures are also evolving. 
In the MHD scale, instabilities, such as Kelvin-Helmholtz (K-H) instability, can cause variations of density and magnetic field at the magnetosphere boundary layer \citep{Hasegawa2004}.
In the plasma scale, plasma waves, such as electromagnetic ion cyclotron (EMIC) waves \citep{Allen2015}, ultra-low-frequency (ULF) waves \citep{Soucek2015,Takahashi2018}, kinetic Alfv\'{e}n waves (KAWs) \citep{Zhao2014}, etc., are widely found in the solar wind, magnetosheath, and magnetosphere.
Turbulence exists at sizes ranging from MHD to plasma scales \citep{He2012,Sahraoui2013,Huang2018}. 
Besides, eruption events from the Sun \citep[e.g., coronal mass ejections, and coronal shocks,][]{Su2015,Su2016} and Earth magetosphere \citep[{e.g., magnetic reconnections,}][]{Huang2012,Takahashi2018,Zhou2019} can lead to variations of density and magnetic field in multiple scales. 
In this work, we evaluate the effect of the OPD noise rooted from the space plasma at the global and MHD scales 
on the detection of GWs for TQ. 

%The rest of the paper is organized as follows. 
%Section \ref{sec:2} gives the formalism that is subsequently used to analyze 
%the acceleration noises of TMs induced by the magnetic field of the space plasma. 
%The MHD model, i.e. Space Weather Modeling Framework (SWMF) \cite{Toth2005}, used 
%to simulate the global magnetosphere of the Earth, and the relative positions of the TQ's orbit  
%in the magnetosphere ($\phi_{s}'$) are discussed in section \ref{sec:3}. 
%In section \ref{sec:4}, based on the magnetic field from the simulation, we calculate 
%the acceleration noises at typical values of $\phi_{s}'$ and the associated amplitude spectral density (ASD), 
%and then evaluate the impact of the acceleration noises from space environment on TQ's sensitivity curve. 
%The paper is concluded in section \ref{sec:5}. 

This paper is organized as follows.
The theory of EM wave propagation in space plasma is briefly summarized in Section \ref{sec:2}. In Section \ref{sec:sim}, we introduce the MHD model, i.e., the Space Weather Modeling Framework \citep[{SWMF};][]{Toth2005}, which is adopted in this work. 
The calculation and results of the OPD noise 
%and the update of the TQ's sensitivity curve that considered the OPD noise 
are presented in Section \ref{results}. 
Section \ref{discussions} discusses the impact of our work on the detection of GWs for TQ. 
Our paper is concluded in Section \ref{conclusions}.

\section{Electromagnetic Wave Propagation in Space Plasma} \label{sec:2}

%The plasma of the Earth magnetosphere and solar wind is considered as cold plasma.
For a train of EM waves with a frequency $f$ (angular frequency $\omega$) propagating in the cold magnetized plasma of the Earth magnetosphere and solar wind, the refractive index $\mu$ can be described by the Appleton-Hartree (A-H) equation \citep{Hutchinson2002}:
\begin{equation}
\label{Eq1}
\mu^2 = 1 - \frac{X}{1 - iZ - \frac{Y_T^2}{2(1-X-iZ)} \pm \sqrt{Y_L^2 + \frac{Y_T^4}{4(1-X-iZ)^2} } }  \,,
\end{equation}
where $i$ represents the imaginary unit, $X$, $Y$, and $Z$ are defined as:
\begin{equation}
\label{Eq2}
X = \frac{\omega_p^2}{\omega^2} \,, ~~ Y = \frac{\omega_B}{\omega} \,, ~~  Z = \frac{\nu}{\omega}  \,.
\end{equation}
Here $\omega_p$, $\omega_B$ and $\nu$ are the plasma frequency, gyrofrequency, 
and electron collision frequency, respectively:
\begin{equation}
\label{Eq3}
\omega_p^2 = \frac{e^2 N}{m \varepsilon_0} \,,  ~~ \omega_B = \frac{eB}{m} \,,
\end{equation}
where $m$ is the electron mass, $e$ is the elementary charge, $N$ is the electron number density, $\varepsilon_0$ is the vacuum electric permittivity, and $B$ is the background magnetic field strength. 
In Equation (\ref{Eq1}), $Y_T = Y\cos\theta$ and $Y_L = Y\sin\theta$, 
where $\theta$ is the angle between the propagation direction of the EM wave and the direction of the background magnetic field.

Since the electron collision frequency $\nu$ of the plasma in the magnetosphere and solar wind are on the order of $10^{-9}$~s$^{-1}$, which are much lower than the frequency of the diode-pumped Nd:YAG laser ($\approx 2.8 \times 10^{14}$ s$^{-1}$) used for TQ, the space plasma can be considered as collisionless. Thus, $Z$ can be ignored.  
%and Equation (\ref{Eq1}) can be simplified as:
%\begin{equation}
%\label{Eq4}
%\mu^2 = 1 - \frac{X}{1 - \frac{Y_T^2}{2(1-X)} \pm \sqrt{Y_L^2 + \frac{Y_T^4}{4(1-X)^2} } }.
%\end{equation}
Besides, take the typical electron number density to be 5 cm$^{-3}$ and the typical magnetic strength to be 5 nT at the geocentric distance of 10$^5$ km, $\omega_p$ and $\omega_B$ in the magnetosphere and solar wind are on the order of $10^{5}$~rad s$^{-1}$ and $10^{3}$~rad s$^{-1}$, respectively, both are much lower than $\omega$.
Therefore, Equation (\ref{Eq1}) can be simplified as,
%	then $(\omega_B/\omega_p)^2 \approx 5\times10^{-5}$. Thus, the terms that contain $Y^2/X$ in can be ignored.
\begin{equation}
\label{Eq5}
\mu^2 = 1 - X  \,.
\end{equation}
%

%Note that $n$ in A-H equation is phase refractive index, 
The group refractive index $\mu_g$ can be deduced as:
\begin{equation}
\label{Eq6}
\mu_g = \frac{\partial (\mu \omega)}{\partial \omega} = \frac{\partial \omega \sqrt{1-X}}{\partial \omega} 
= \frac{2 \omega}{2 \sqrt{\omega^2 - \omega_p^2}} = \frac{1}{\sqrt{1-X}} = \frac{1}{\mu} \approx 1 + \frac{X}{2} = 1 + \frac{K N}{2f^2} \,,
\end{equation}
where $K = e^2/(4\pi^2 m \varepsilon_0) = 80.6~\rm m^3~s^{-2}$. 

The time ($\tau$) that takes EM waves propagating a distance of $L$ in space plasma is:
\begin{equation}\label{Eq7}
\tau = \int_{L} \frac{\mathrm{d} s}{v_g} = \int_{L} \frac{\mathrm{d} s}{c/\mu_g}  \,,
\end{equation}
%where $L$ is the propagation distance of EM wave, and 
where $c$ is the speed of light in vacuum, $L \approx 1.7 \times 10^8$ m for TQ. 
The time delay ($\Delta \tau$) relative to the vacuum case is:  %caused by the EM waves propagation in space plasma is:
\begin{equation}
\label{Eq8}
\Delta \tau = \frac{1}{c} \int_{L} (1 + \frac{K N}{2f^2}) \mathrm{d} s - \frac{L}{c} = \frac{K}{2 c f^2} \int_{L} N \mathrm{d} s  \,.
\end{equation}
Here, $\int_{L} N \mathrm{d} s$ is called TEC.  According to Equation (\ref{Eq8}), the OPD can be calculated as: 
\begin{equation}
\label{Eq9}
\Delta l = c \Delta \tau = \frac{K}{2 f^2} \int_{L} N \mathrm{d} s  \,.
\end{equation}
Equation (\ref{Eq9}) shows that the OPD noise $\Delta l$ of a single arm between two satellites is determined by the integrated electron number density along the laser link.

\section{MHD Simulation}\label{sec:sim}

According to Section \ref{sec:2}, in order to study the OPD noise for TQ, we need to obtain the distributions of the electron number density in the vicinity of the laser links in Figure~\ref{fig2} (see also Figure 1 in \citet{Su2020A}),
which requires the global MHD simulations of the Earth magnetosphere.
In this work, we adopt the Space Weather Modeling Framework (SWMF) to simulate the interaction 
between the solar wind and the Earth magnetosphere \citep{Toth2005}. 
SWMF has been thoroughly validated in the study of the Earth magnetosphere \citep{Zhang2007,Welling2010,Dimmock2013}, and it has been used widely \citep{Lv2015,Wang2016,Takahashi2018}. 
The simulation can be requested on the Community Coordinated Modeling Center (CCMC),
which is done by the SWMF/Block-Adaptive-Tree-Solarwind-Roe-Upwind-Scheme (BATSRUS). 

\begin{figure}[ht!]
%	\figurenum{1}
	\plotone{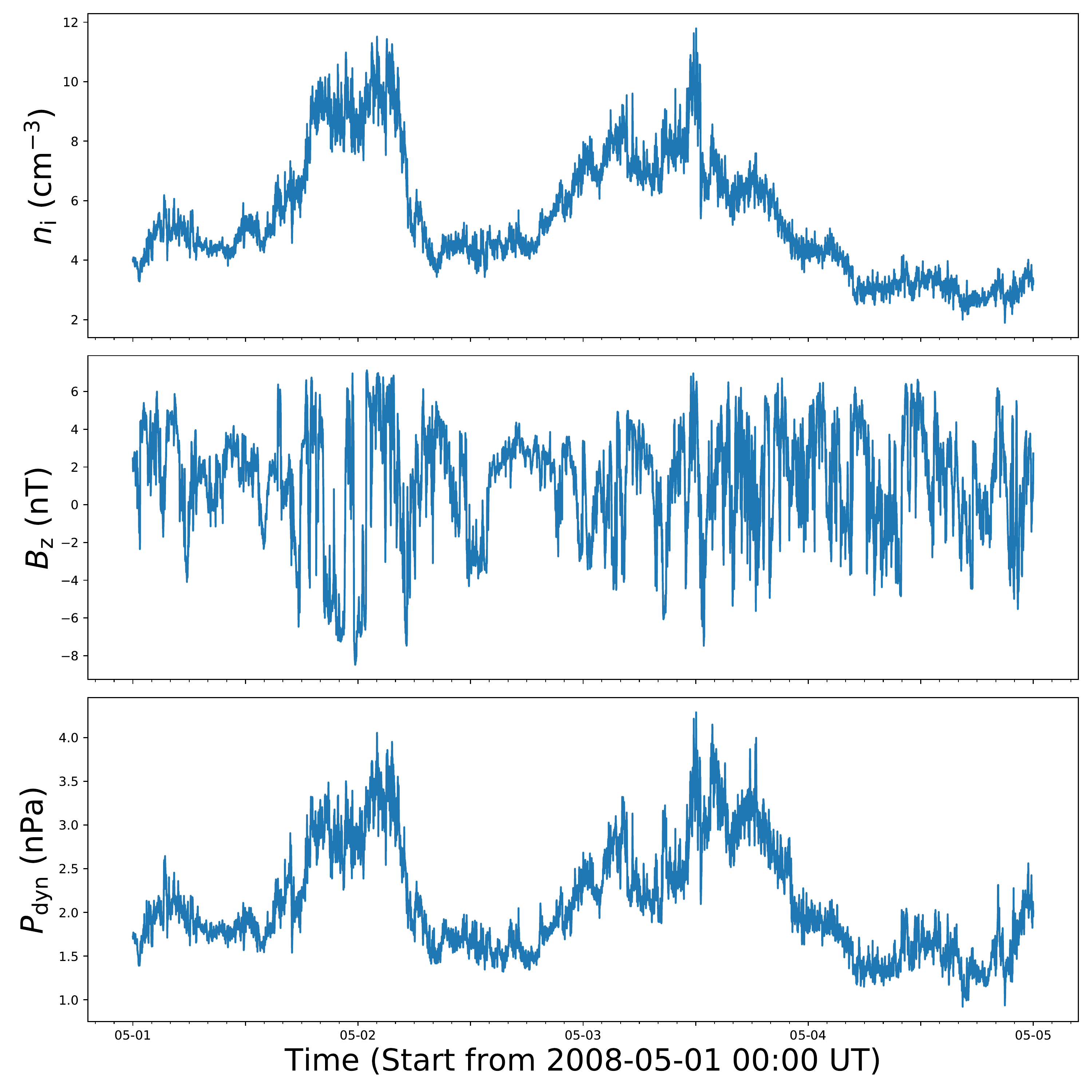}
	\caption{The input parameters observed by the ACE spacecraft which include the ion number density $n_i$ (top panel), the space magnetic field $B_{\rm z}$ in the GSM coordinate (middle panel). The solar wind dynamic pressure $P_{\rm dyn}$ (bottom panel) is derived from the observations of the ACE. }
	\label{fig1}
\end{figure}

The real time solar wind parameters observed by the Advanced Composition Explorer \citep[{ACE};][]{Stone1998} are taken as the simulation inputs, which include the ion number density $n_i$, $z$ component of magnetic field $B_z$, and solar wind dynamic pressure $P_{\rm dyn}$ as illustrated by Figure \ref{fig1}. The time range of the inputs is from 2008-05-01 00:00 UT to 2008-05-04 24:00 UT with a temporal resolution of 1 min. The input data are the same as in \citet{Su2020A}.
The ranges of the Geocentric Solar Magnetospheric (GSM) coordinates in the simulation domain are $-250R_{E} <x< 33R_{E}$ ($R_{E}$, the radius of the Earth), $\left| y \right|$ and $\left| z \right| < 48R_{E}$, which contain the solar wind in the interplanetary space, the bow shock, magnetopause and magnetotail of the Earth. 
In the region where $\left| x \right|, \left| y \right|, \left| z \right|< 20R_{E}$, the vicinity of the dayside magnetopause and the near-tail has the finest resolution of 0.25$R_{E}$, the resolution of the rest region is 0.5$R_{E}$. 
The output parameters of the simulation contain the magnetic field ($B_{x}$, $B_{y}$, $B_{z}$), the plasma parameters (e.g., bulk flow velocity $v_{x}$, $v_{y}$, $v_{z}$, number density of ions $n_i$, pressure $P$), and electric current ($J_{x}$, $J_{y}$, $J_{z}$). 
The output parameters in the GSM coordinates are converted to the Geocentric Solar Ecliptic (GSE) coordinates in the following calculation. 
%In the simulation, the number density of eletrons $n_e$ and ions $n_i$ set to be equal,
%in addition, $n_e$ is approximately equal $n_i$ in the solar wind and magnetosphere in the observation results \citep{Zhang2007,Welling2010}, we take the output of $n_i$ instead of $n_e$ in the calculation of the OPD noise. 
Generally, the plasma in the solar wind and magnetosphere is quasi-neutral at the MHD scale, 
and the number densities of electrons and ions $n_i$ are approximately equal. 
This has been confirmed by several observations \citep{Zhang2007, Welling2010}. 
Therefore, we simply use $n_i$ outputted from the simulation as electrons number density in the calculation of the OPD noise. 
%%%%%

%orbit plane

In the GSE coordinates, we define the intersection angle between the Sun-Earth vector and the projection of the normal of the detector plane on the ecliptic plane as $\phi_{s}$, which shows an annual variation from 0$^\circ$ to 360$^\circ$ \citep{Su2020A} 
and is equal to 120.5$^\circ$ at the spring equinox \citep{Hu2018}. 
In order to describe the relative position of the geometric structure of the Earth magnetosphere and the TQ's constellation  conveniently,  $\phi_{s}$ is transformed to its corresponding acute angle $\varphi_{s}$ hereafter. 
%$\varphi_{s}$ ranges from 0$^{\circ}$ to 45$^{\circ}$ in TQ's observation time intervals, while it ranges from 45$^{\circ}$ to 90$^{\circ}$ in non-observation time intervals. %(see Fig. 4 and the associated text in \citet{Su2020A} for details). 
%More details can be found in \citet{Su2020A}.
$\varphi_s$ can be approximately regarded as a constant during one orbit period of the TQ satellite around the Earth \citep[3.65 days, ][]{Su2020A}. 
In the following sections, we focus on the OPD noises at four typical positions 
with $\varphi_{s} =$  0$^{\circ}$, 30$^{\circ}$, 60$^{\circ}$, and 90$^{\circ}$.

\section{Results}\label{results}

\subsection{Laser links in magnetosphere and OPD noise}

\begin{figure}[ht!]
%	\figurenum{2}
	\plotone{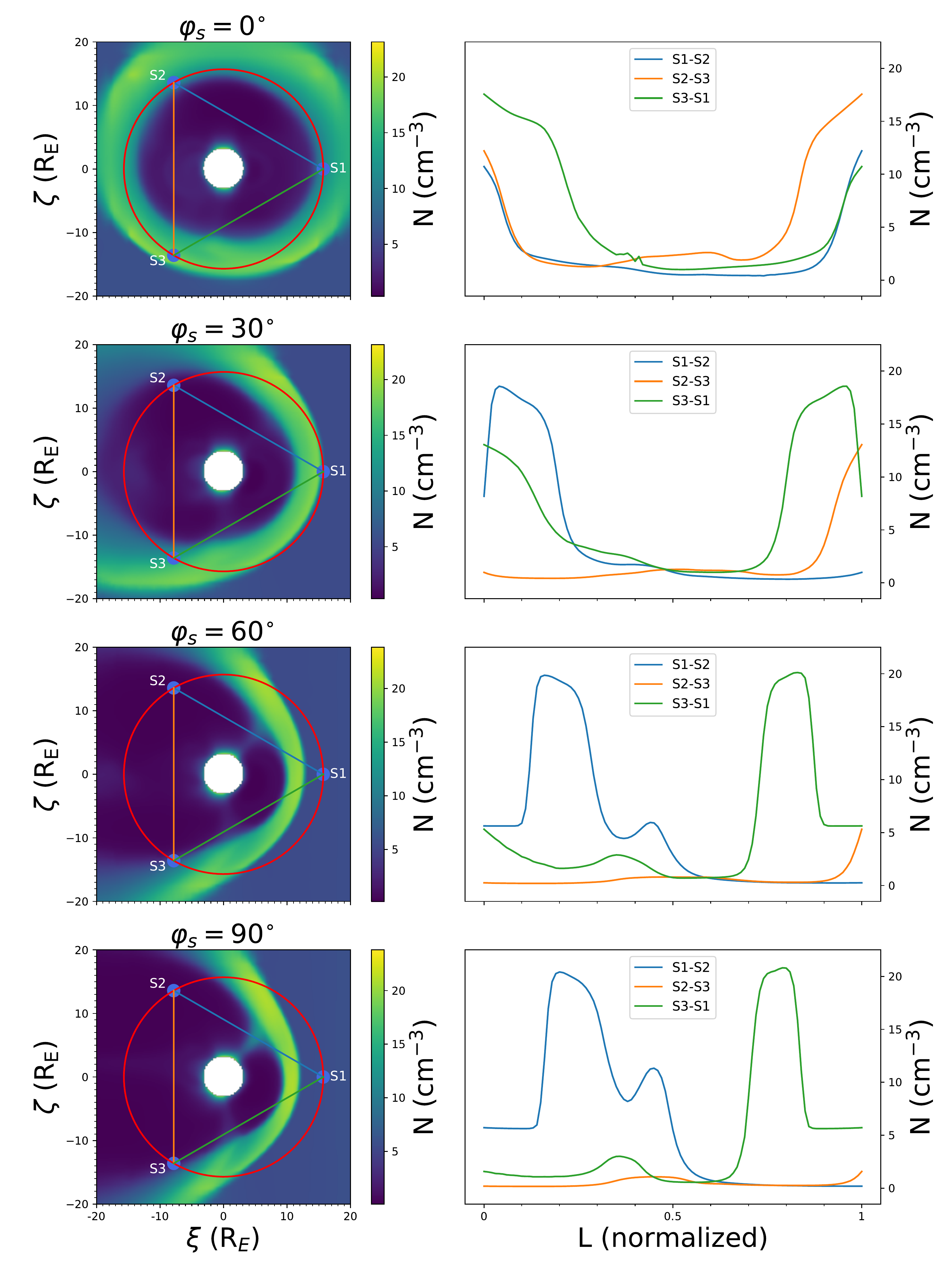}
	\caption{The left column displays the electron number density distributions (color scale) on the orbit planes with $\varphi_s = 0^{\circ}, 30^{\circ}, 60^{\circ}, 90^{\circ}$. $\xi$ is the intersection line between the orbit plane and the ecliptic, $\zeta$ is perpendicular to the ecliptic, the red circle is the orbit of TQ's satellites. The laser link S1 -- S2, S2 -- S3, and S3 -- S1 are represented by blue, orange, and green lines, respectively. The right panels are the distributions of the electron number densities along these three laser beams. Different colors correspond to the laser beams with the same colors in the left column. }
	\label{fig2}
\end{figure}

Taking the simulation at 2008-05-03 20:00 UT as an example, the electron number density distributions 
on the detector's planes (at $\varphi_s = 0^{\circ}, 30^{\circ}, 60^{\circ}, 90^{\circ}$) are shown on the left columns of Figure \ref{fig2}, in which $\xi$ is the intersection line between the orbit plane and the ecliptic plane, and $\zeta$ is along the intersection of the detector's plane and a plane perpendicular to $\xi$. 
$\zeta$ is approximately vertical to the ecliptic plane since the angle between the ecliptic plane and the normal of the detector's plane is only 4.7$^{\circ}$ \citep{Hu2018}. 
%The areas with high number density value in Fig. \ref{fig2} is the Earth magnetosheath, in the different orbit planes. The magnetosheath is the downstream of the Earth bow shock, the boundary of the magnetosheath on the sunside and Earthside are the bow shock and magnetopause, respectively. 
The Earth magnetosheath is the downstream of the bow shock, therefore its electron number density is higher than the ones in the solar wind and the magnetosphere. As shown in Figure \ref{fig2}, the boundary of the magnetosheath on the sunside and earthside are the bow shock and magnetopause, respectively. 
The geometric structures of the magentosphere on the four detector's planes are different.
For $\varphi_s$ = 90$^{\circ}$, the nose of the bow shock is located at $\xi \approx 10~{R_E}$ measured from the geocenter. 
For $\varphi_s$ = 0$^{\circ}$, the magnetopause and bow shock are approximately circular 
and they are located at $\approx 15~{R_E}$ and $\gtrsim 20~{R_E}$, respectively.

%The schematic view of TQ's laser links are shown in Fig. 1 of \citet{Su2020A}, 
With the time-varying positions of three TQ's satellites (S1, S2, and S3), the laser links can be obtained. In Figure \ref{fig2}, the laser links S1--S2, S2--S3, and S3--S1 are represented as blue, orange, and green lines, respectively. Note that the initial position of S1 is located at $\xi = 15.7~\rm{R_E}$ and $\zeta = 0~\rm{R_E}$. 
For $\varphi_s$ = 60$^{\circ}$ and 90$^{\circ}$, S1--S2 and S3--S1 will pass through the solar wind, bow shock, magnetosheath, magnetopause and magnetosphere; While S2--S3 that passes through the magnetotail is almost enclosed in the magnetosphere. 
We obtain the number density distributions along these three laser links, shown in the corresponding colors in the right column of Figure~\ref{fig2}, by interpolating the values of number densities on the grid of the simulation domain. 
The number density characteristics of the regions, such as the solar wind (moderate), magnetosheath (high), and magnetosphere (low), are also revealed here. 
%As the satellites of TQ move, the laser links will also move in the detector's plane.
%The laser links sweep across the annulus formed by the inscribed circle and the circumcircle (red circle) of the detector's triangle as the satellites orbiting around the geocenter. 

%In space, with the movement of TQ's satellites, 
% %in accordance with satellite movement, 
%which leads to the changes of the number density distributions along the laser beams; 
%%In temporal, with the variation of the inputs, 
%In time, the geometrical shapes and the number densities of the bow shock, magnetosheath and magnetopause will evolve, thus, the number density distributions along each laser link will vary.
%%The most important factor of the inputs affecting the geometry shape and number density of the magnetosphere is the solar wind dynamic pressure $P_{dyn}$. The stronger $P_{dyn}$ is, the stronger the magnetosphere is compressed, and the larger the number density in the magnetosheath. Otherwise, the opposite.
%%Meanwhile, considering that the OPD noise is proportional to the integration of the electron number density, the OPD noise is positively correlated with $P_{dyn}$.
%%The weaker the solar wind dynamic pressure, the opposite.

\begin{figure}[ht!]
%	\figurenum{3}
	\plotone{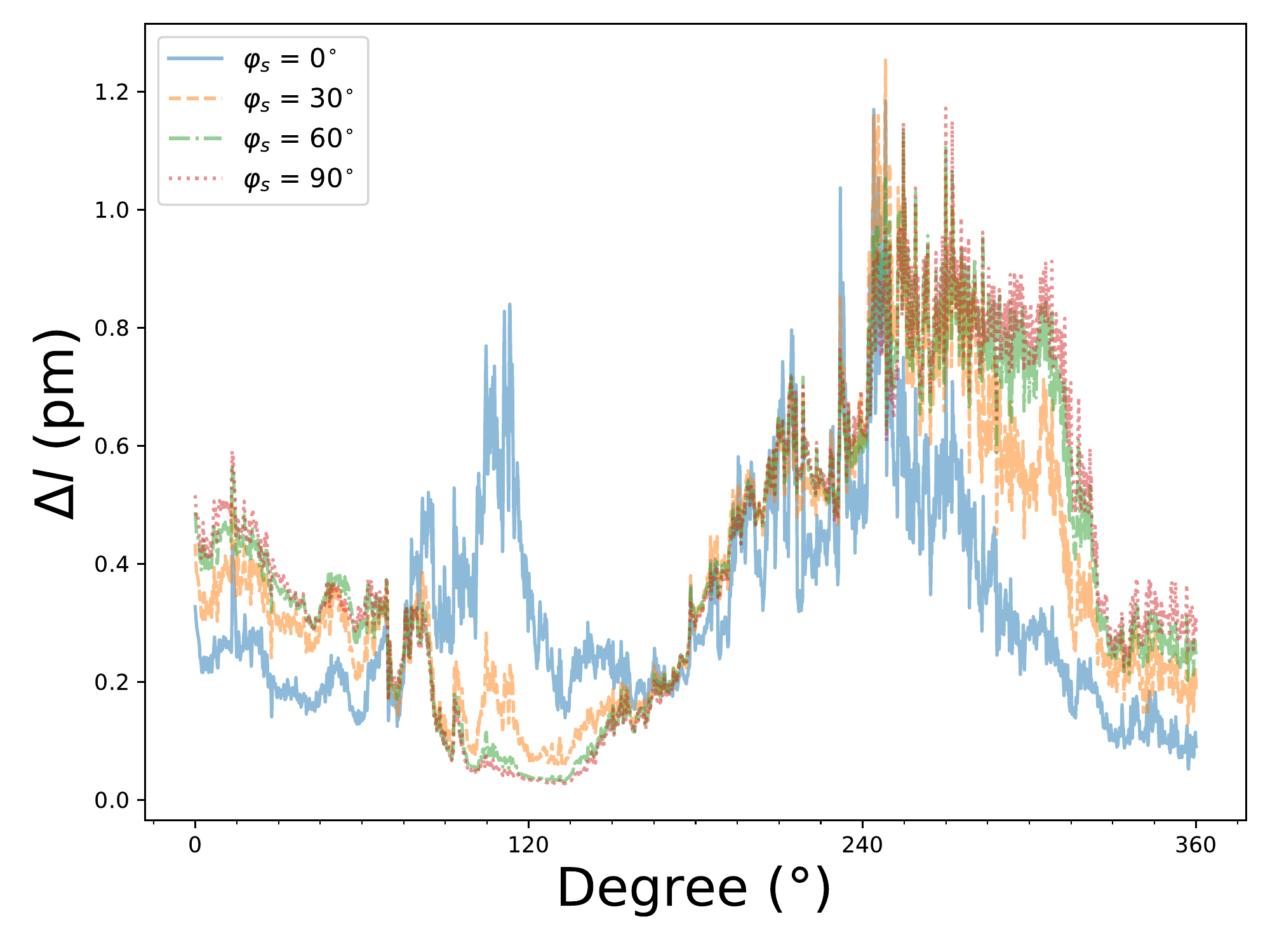}
	\caption{The OPD noises for a single link during a 3.65-day full-period orbit around the Earth on the orbit planes with $\varphi_s = 0^{\circ}$, $30^{\circ}$, $60^{\circ}$, and $90^{\circ}$.}
	\label{figDistribS}
\end{figure}

The laser links sweep across the annulus formed by the inscribed circle and the circumcircle (red circle) of the detector's triangle as the satellites orbiting around the geocenter. 
We calculate the OPD noise $\Delta l$ based on Equation (\ref{Eq9}) and the satellites' orbits. 
The integration of the number density along each laser beam shows spatial and temporal variations due to  
the changes of the positions and directions of the laser beams in the magnetosphere and 
the evolution of the geometrical shapes and the number densities of the structures. 
During one revolution of the satellites around the Earth, the time series of $\Delta l$ of S1--S2 
for $\varphi_s = 0^{\circ}$, $30^{\circ}$, $60^{\circ}$, $90^{\circ}$ are shown in Figure \ref{figDistribS}. 
For $\varphi_s$ = 0$^{\circ}$, the correlation coefficient between the time series of $P_{\rm dyn}$ and $\Delta l$ is 0.83, which is about two to three times larger than the ones for $\varphi_s = 30^{\circ}$ (0.48), $60^{\circ}$ (0.33), and $90^{\circ}$ (0.28).  
For $\varphi_s$ = 30$^{\circ}$, 60$^{\circ}$, 90$^{\circ}$, the amplitude of the OPD noise 
reaches 1.2 pm at the position when the laser beam passes through the magnetosheath 
on the dayside (around 300$^{\circ}$ in Figure~\ref{figDistribS}), while $\Delta l$ is only about 
0.05 pm at the position where the laser beam passes through the magnetotail on 
the nightside (around 120$^{\circ}$ in Figure~\ref{figDistribS}).
%This difference is mainly due to the number density in the magnetosheath is much higher than that in the magnetotail. 
%For $\varphi_s$ = 0$^{\circ}$, since quasi-axisymmetric structure of the magnetosheath along the $−x$ axis in the GSE coordinates, $\Delta l$ are quasi-constant in spatial, however, the variation of $P_{dyn}$ in temporal can lead to the variation of d$l$. 
%And since quasi-axisymmetric structure of the magnetosheath along the $−x$ axis in the GSE coordinates for $\varphi_s = 0^{\circ}$, the OPD noises are quasi-constant in spatial. 
These results indicate that the variation of $\Delta l$ for $\varphi_s = 0^{\circ}$ is mainly 
due to the evolution of $P_{\rm dyn}$ in time,
whereas the variations of $\Delta l$ for $\varphi_s$ = 30$^{\circ}$, 60$^{\circ}$, 90$^{\circ}$ 
are mainly due to the fact that the number density in the magnetosheath is much higher than that in the magnetotail. 

%With the electron number density along the laser link, we can calculate the OPD (d$l$) noise at one moment by Equation (\ref{Eq9}). Here, $f$ in Equation (\ref{Eq9}) is taken as the \textcolor{blue}{Nd:YAD} laser frequency of TQ 2.8 $\times 10^{14}$ Hz. In a period of satellite around the Earth (one period), 3.65 day, we can obtain the OPD noise of S1--S2 link every 60 s (time resolution of the simulation). In this way, we get the distribution of the OPD noise in one period for $\varphi_s = 0^{\circ}, 30^{\circ}, 60^{\circ}, 90^{\circ}$, and the results are shown as blue curves in Fig. \ref{fig3}. In Fig. {\ref{fig3}}, we find that the maximum value of the OPD noise can reach 1.2 pm, which is larger than the detection accuracy of TQ (1 $\rm pm/\sqrt{Hz}$). For $\varphi_s$ = 30$^{\circ}$, 60$^{\circ}$, 90$^{\circ}$, we can see that when the laser link passes through the magnetosheath on the dayside, the OPD noise is relatively large (the corresponding Degree around 300$^{\circ}$ in Fig. \ref{fig3}); when the laser link passes through the magnetotail on the nightside, the OPD noise is relatively small. This is because the electron number density of the magnetosheath is significantly higher than that of the magnetotail. 
%This phenomenon indicates that the OPD noise is related to the position of the laser link, and the noise is nonstational. Besides, $P_{dyn}$ is a significant parameter to indicate the electron number density of the magnetosheath, the larger $P_{dyn}$ is, the larger $n_i$ in magnetosheath is.

\subsection{OPD noise for single link and Michelson interferometer }
%\label{result2}

\begin{figure}[ht!]
%	\figurenum{4}
	\plotone{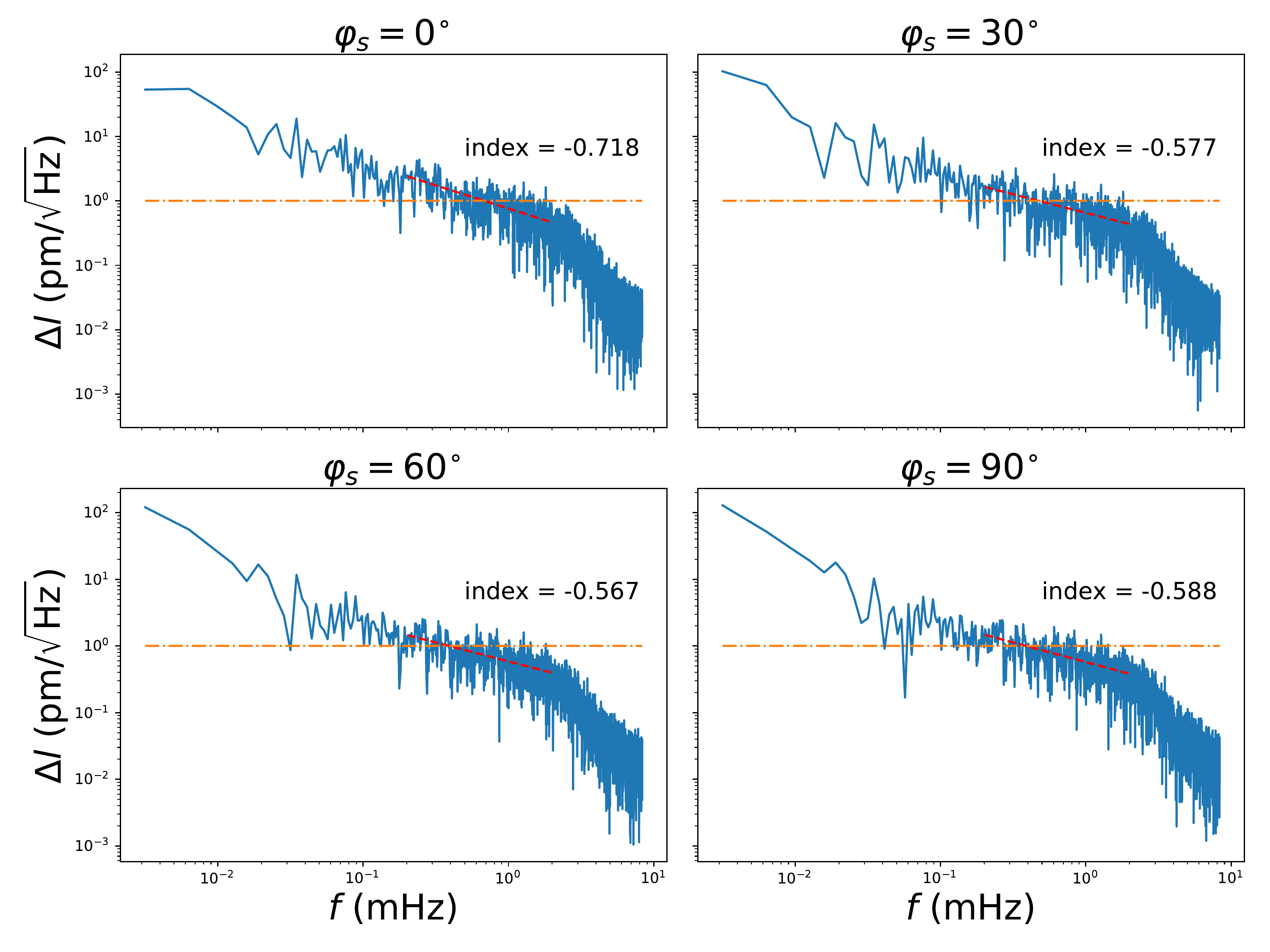}
	\caption{The ASDs of the OPD noises for a single link on the orbit planes with $\varphi_s = 0^{\circ}$, $30^{\circ}$, $60^{\circ}$, and $90^{\circ}$. The orange dashed line is the proposed displacement measurement accuracy of TQ (1 pm/$\rm{\sqrt{Hz}}$) \citep{Luo2016}. Red dashed lines are the best-fit spectra of the ASDs. }
	\label{figASDS}
\end{figure}

Figure \ref{figASDS} shows the amplitude spectral densities (ASDs) of the time series of $\Delta l$ for the single links. % S1--S2. 
Here, we have used Savitzky-Golay filter \citep{SavGol1964} to smooth the ASDs %of d$l$ 
before fitting them by a single power law function. 
Note that the finest spatial resolution of the simulation is 0.25 $R_E$ and the speed of TQ satellites is about 2 km s$^{-1}$, 
it takes each satellite about 800 s to move between two grid points. 
So that the ASDs of the OPD noise at range of $f>1/800$ Hz can be underestimated. 
In fact, this underestimation has been shown in Figure \ref{figASDS}, where there is a knee point 
at $f\approx1/800$~Hz and the spectra become steeper when $f \gtrsim 1/800$~Hz. 
Here, only the ASDs of the OPD noise at range of $f \lesssim 1/800$ Hz are used in the fitting of the spectral index. 
The best-fit spectral indices for $\varphi_s = 0^{\circ},~30^{\circ},~60^{\circ}$ and $90^{\circ}$, shown as the red dashed lines, 
are -0.718, -0.577, -0.567 and -0.588, respectively. 
The corresponding spectral amplitudes at 1 mHz read 0.760, 0.651, 0.587 and 0.573 pm/$\sqrt{{\rm Hz}}$, respectively. 

Michelson-type interferometer sketched in Figure \ref{figTDIpath}a has been used as the fiducial data combination 
to study the science potential and data analysis for space-borne detectors \citep{Feng2019, 2019PhRvD.100d3003W,PhysRevD.101.103027, PhysRevD.102.063021}. 
Its response and sensitivity to arbitrary incoming GWs for TQ have been studied in \citet{Hu2018}. 
We denote the OPD noise that is produced during the propagation of the EW wave 
sent from spacecraft $i$ and received by satellite $j$ as $\Delta l_{ij}$ \citep{Prince2002}. 
For a Michelson-type interferometer centered on S1, the OPD noise of two interferometer arms (S1--S2 and S1--S3) 
are $\Delta l_{12}$, $\Delta l_{21}$, $\Delta l_{13}$ and $\Delta l_{31}$. 
From Equation (\ref{Eq9}), $\Delta l_{12} = K/2 f^2 \int_{S1}^{S2} N \mathrm{d} s$, 
and similarly for $\Delta l_{21}$, $\Delta l_{13}$ and $\Delta l_{31}$. 
Since $\int_{i}^{j} N \mathrm{d} s =  \int_{j}^{i} N \mathrm{d} s$, and the light propagation time between a pair of satellites ($\approx 0.6$~s) is much less than the temporal resolution of our simulation (60 s), we set $\Delta l_{12}$ = $\Delta l_{21}$ and $\Delta l_{13}$ = $\Delta l_{31}$. 
Therefore, the OPD noise for a Michelson-type combinations can be written as follows,
\begin{equation}
\label{Eq10}
\Delta l = 2 (\Delta l_{12} - \Delta l_{13} )  \,.
\end{equation}
From Equation (\ref{Eq10}), we can calculate the time series of $\Delta l$ for a Michelson combination during one revolution of the satellites around the Earth.
From Figure~\ref{figDistribM}, we can see that the maxima of $|\Delta l|$ for the Michelson combination is about 3 pm. 
The typical amplitudes are magnified due to four combinations of the OPD noises of the single links. 

Shown in Figure \ref{figASDM} are the ASDs of $\Delta l$ for the Michelson interferometer 
for $\varphi_s = 0^{\circ}$, $30^{\circ}$, $60^{\circ}$, and $90^{\circ}$. 
Similarly, % to the fitting for the OPD noises of the single link, 
we fit the spectral profiles with power-law functions for the Michelson interferometer, which are shown as the red dashed lines in Figure \ref{figASDM}. 
The best-fit spectral indices for $\varphi_s = 0^{\circ},~30^{\circ},~60^{\circ}$, and $90^{\circ}$ 
are -0.887, -0.544, -0.626 and -0.683. 
The corresponding spectral amplitudes at 1 mHz read 0.752, 1.167, 1.400 and 1.512 pm/$\sqrt{{\rm Hz}}$, respectively.

\subsection{OPD noise for TDI combinations}

\begin{figure}[ht!]
%	\figurenum{4}
	\plotone{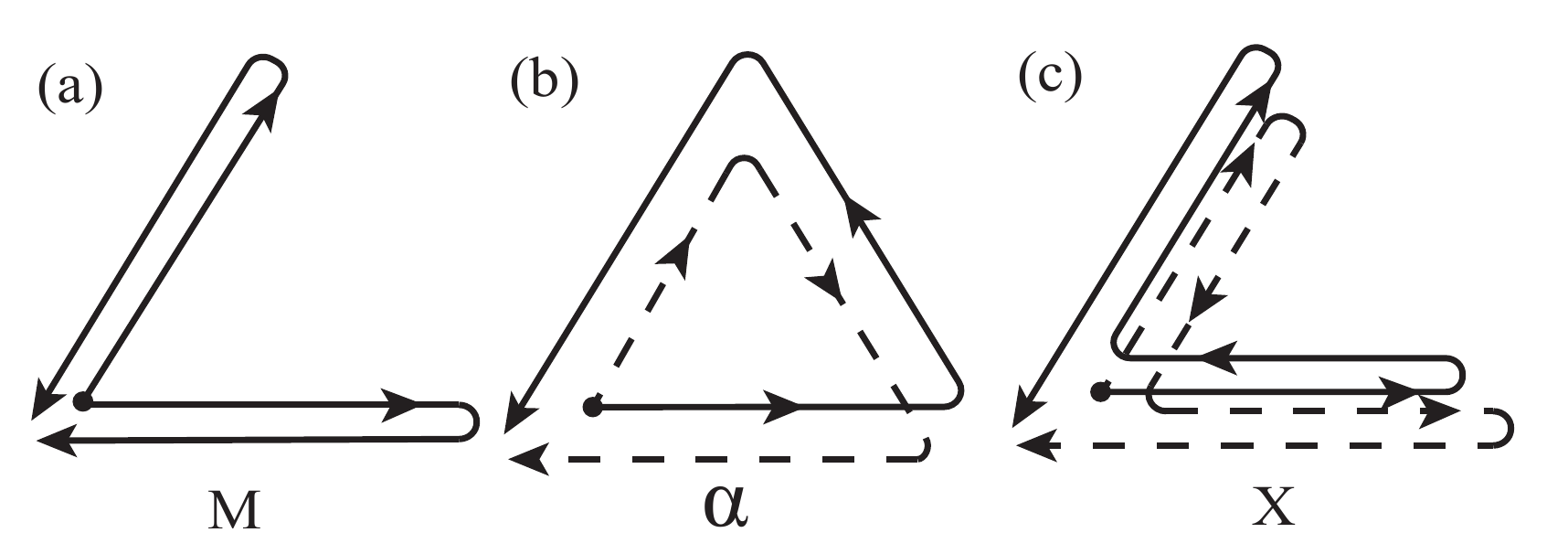}
	\caption{Schematic diagrams of Michelson, $\alpha$, and $X$ combinations. }
	\label{figTDIpath}
\end{figure}

\begin{figure}[ht!]
%	\figurenum{5}
	\plotone{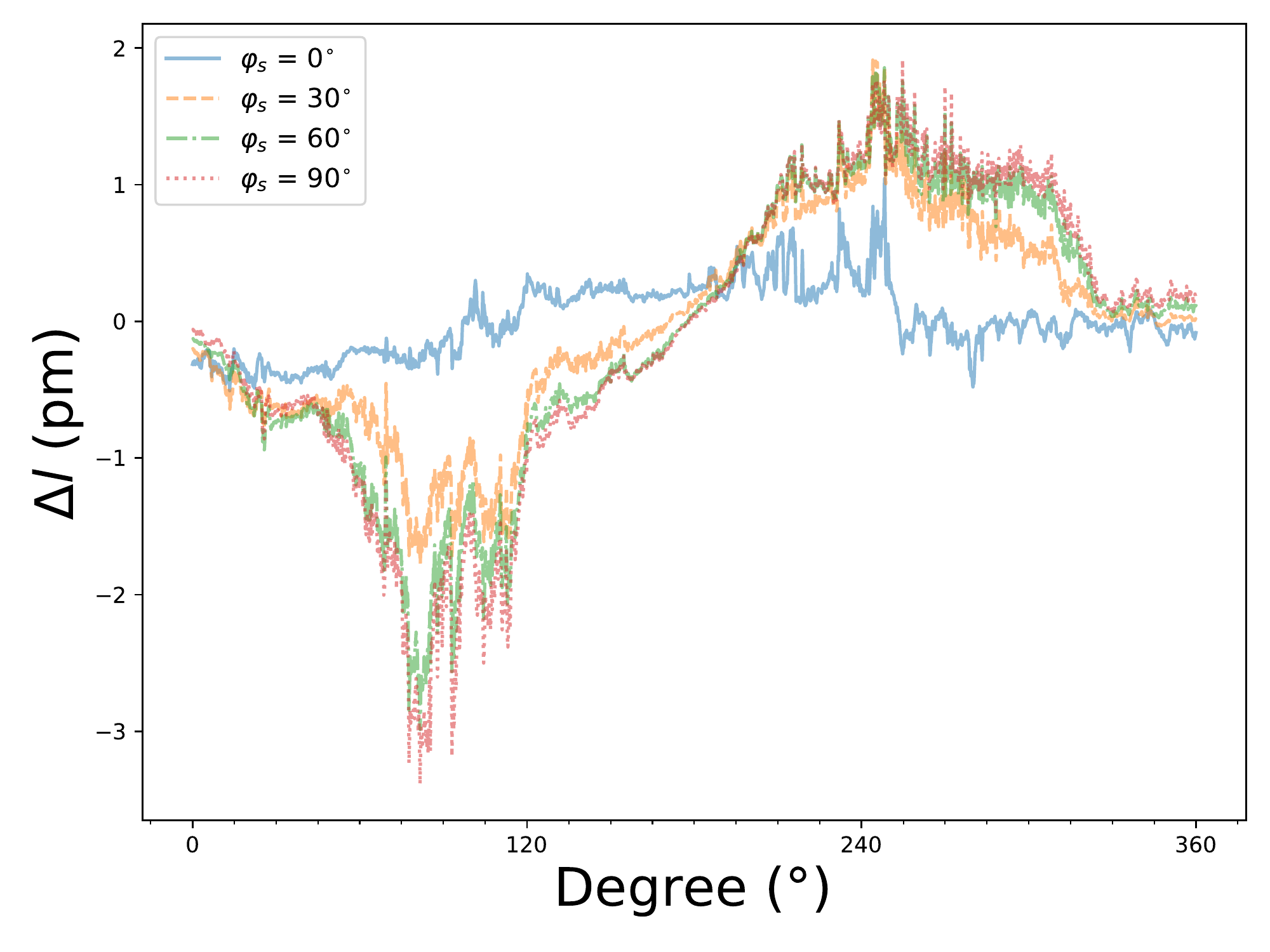}
	\caption{The OPD noises for a Michelson combination  
		during a 3.65-day full-period orbit around the Earth. The blue, orange, green, and red curves are the OPD noises
		on the detector's planes with $\varphi_s = 0^{\circ}, 30^{\circ}, 60^{\circ}$, and $90^{\circ}$.}
	\label{figDistribM}
\end{figure}

\begin{figure}[ht!]
%	\figurenum{6}
	\plotone{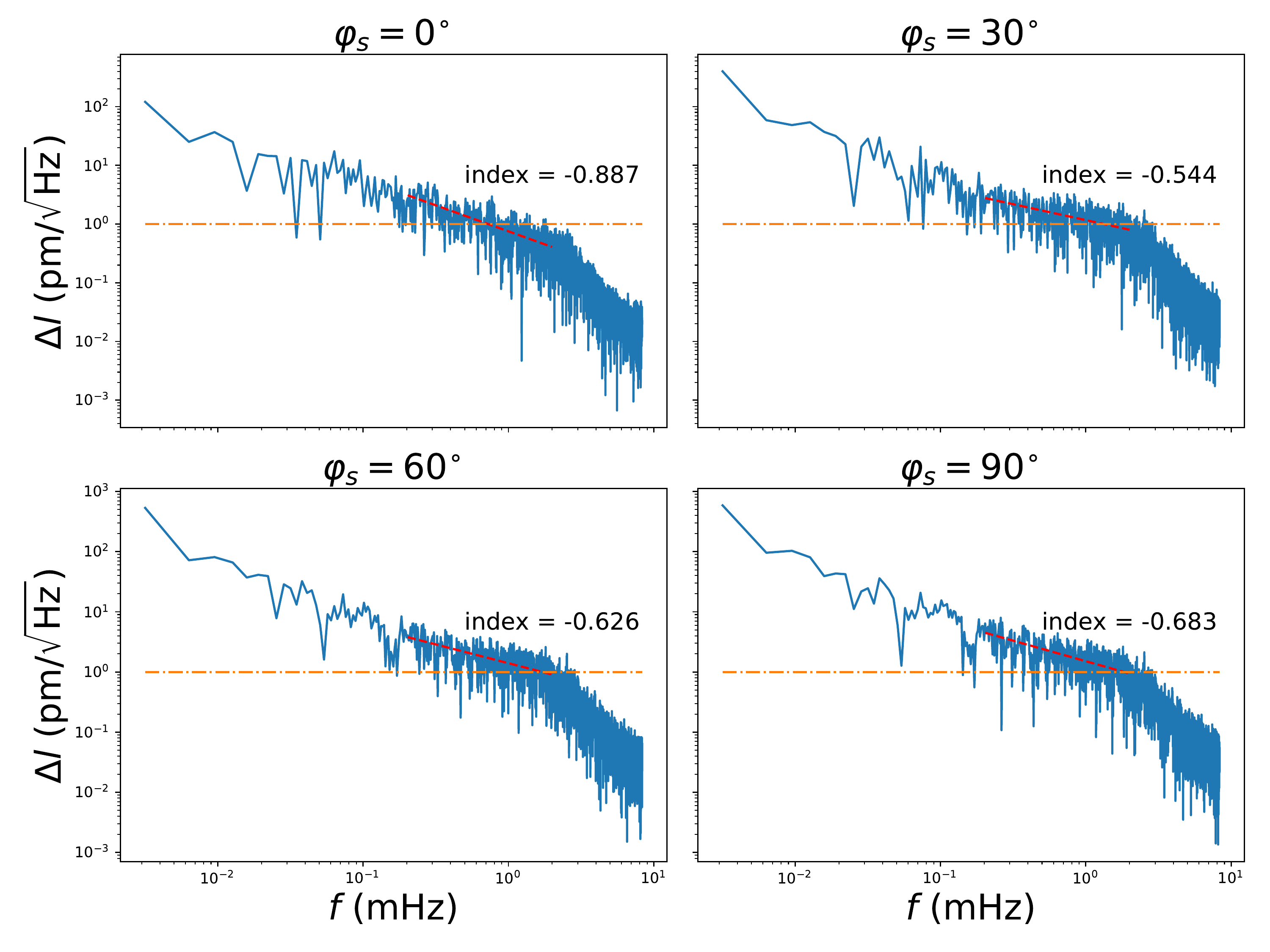}
	\caption{The ASDs of the OPD noises for the Michelson combination on the orbit planes with $\varphi_s = 0^{\circ}$, $30^{\circ}$, $60^{\circ}$, and $90^{\circ}$. The orange dashed line is the proposed displacement measurement accuracy of TQ (1 pm/$\rm{\sqrt{Hz}}$) \citep{Luo2016}. Red dashed lines are the best-fit spectra of the ASDs.}
	\label{figASDM}
\end{figure}

In order to eliminate the otherwise overwhelming laser phase noise, 
the time delay interferometry (TDI) has been devised for the data processing of  
space-borne interferometric GW detectors \citep{Armstrong1999,Estabrook2000,Tinto2014}. 
There are various data combinations for the TDI \citep{Tinto2014}. 
In this work, we focus on the $\alpha$ and $X$ data combinations, 
shown in Figure \ref{figTDIpath}b and \ref{figTDIpath}c, 
as the typical examples of the six-pulse and eight-pulse combinations of the first-generation TDI, respectively. 
%The most typical TDI combination are $\alpha$ and $X$ combinations \citep{Armstrong1999,Estabrook2000,Tinto2014}, and the schematic diagrams of $\alpha$ and $X$ combinations are shown in Fig. \ref{figTDIpath}b and \ref{figTDIpath}c. 

Consider the OPD noise accumulated along the laser propagation in space plasma, the phase fluctuation of the laser that 
is sent from satellite $i$ and received by satellite $j$ can be expressed as follows, 
\begin{equation}
\label{Eq11}
\Phi_{ij}(t) = \phi_{ij}(t) + h_{ij}(t)+n_{ij}(t) + s_{ij}(t)  \,,
\end{equation}
where $\phi_{ij}(t)$ is the laser phase noise to be canceled by TDI, $h_{ij}(t)$ is the GW signal, and $n_{ij}(t)$ is the total nonlaser phase noise \citep{Hellings2001}. $s_{ij}(t)$ is the phase noise 
associated with the OPD noise $\Delta l_{ij}(t)$, 
\begin{equation}
\label{Eq12}
s_{ij}(t) = 2\pi \Delta l_{ij}(t)/\lambda  \,, 
\end{equation}
where $\lambda$ = 1064 nm for the Nd:YAG laser used by TQ. 

Set $L_{ij}$ as the distance between satellites $i$ and $j$ and $c$ = 1 hereafter, 
the total phase noise due to space plasma for the $\alpha$ combination, $s_{\alpha}$, can be written as,
\begin{equation}
\label{Eq13}
s_{\alpha} = s_{12}(t - L_{23}-L_{31})+s_{23}(t-L_{31})+s_{31}(t) - s_{13}(t - L_{32}-L_{21})-s_{32}(t-L_{21})-s_{21}(t)  \,.
\end{equation}
Similarly, for the $X$ combination, $s_{X}$ can be written as,
\begin{equation}
\label{Eq14}
\begin{aligned}
s_{X} = s_{12}(t-L_{21}-L_{13}-L_{31})+s_{21}(t-L_{13}-L_{31})+s_{13}(t-L_{31})+s_{31}(t) \\
-s_{13}(t-L_{31}-L_{12}-L_{21}) -s_{31}(t-L_{12}-L_{21}) -s_{12}(t-L_{21}) -s_{21}(t)  \,. 
\end{aligned}
\end{equation}
For the nearly equilateral triangular constellation of TQ, $L_{ij} = L \approx 0.6$~s. %all of $L_{ij}$ are equal. 
Note that $L$ is much smaller than the temporal resolution of the MHD simulation, $\Delta t$ = 60 s. 
As mentioned in Section 4.2, $\int^{i}_{j}N\mathrm{d}s$ =$\int^{j}_{i}N\mathrm{d}s$, so that $s_{ij}(t) = s_{ji}(t)$. 
%Since $\Delta l$ is dominant by the integral of the number density along $i$ and $j$, and $\int^{i}_{j}N\mathrm{d}s$ =$\int^{j}_{i}N\mathrm{d}s$, so that $s_{ij}$ is equal to $s_{ji}$. 
Thus, $s_{\alpha}$ is reduced to
\begin{equation}
\label{Eq15}
\begin{aligned}
s_{\alpha} = s_{12}(t - 2L)+s_{31}(t) - s_{12}(t)-s_{31}(t - 2L)  \,.
\end{aligned}
\end{equation}
And $s_{X}$ is reduced to
\begin{equation}
\label{Eq16}
\begin{aligned}
s_{X} = s_{12}(t-3L)+s_{12}(t-2L)+s_{31}(t-L)+s_{31}(t)-s_{12}(t-L)-s_{12}(t)-s_{31}(t-3L)-s_{31}(t-2L)  \,.
\end{aligned}
\end{equation}

As in the single link case, we can obtain $s_{ij}(t)$ for every 60~s.  %, and we calculate the variation of $s_{ij}$ during 0.6 s by linear interpolation. } 
$s_{ij}$ at delayed times can be obtained by linear interpolation,  
i.e., $s_{ij}(t -  \delta t) =  s_{ij}(t) + (s_{ij}(t -  \Delta t)- s_{ij}(t))(\delta t/\Delta t)$.
In this way, Equations (\ref{Eq15}) and (\ref{Eq16}) can be modified as follows, 
\begin{equation}
\label{Eq17}
\begin{aligned}
s_{\alpha} = (s_{12}(t - \Delta t)- s_{12}(t) )\frac{2L}{\Delta t} - (s_{31}(t - \Delta t) - s_{31}(t))\frac{2L}{\Delta t} \,,
\end{aligned}
\end{equation}
\begin{equation}
\label{Eq18}
\begin{aligned}
s_{X} =  2s_{\alpha}  \,.
\end{aligned}
\end{equation}
%\begin{equation}
%\label{Eq18}
%\begin{aligned}
%s_{X} = (s_{12}(t-3\Delta t)+s_{12}(t-2\Delta t) -s_{12}(t-\Delta t)-s_{12}(t))\frac{L}{\Delta t} \\
%+(s_{31}(t)+s_{31}(t-\Delta t)-s_{31}(t-2\Delta t)-s_{31}(t-3\Delta t))\frac{L}{\Delta t}  \,.
%\end{aligned}
%\end{equation}

\begin{figure}[ht]
%	\figurenum{7}
	\centering
	\begin{minipage}[b]{\textwidth}
		\centering
		\begin{minipage}[b]{0.419\textwidth}
			\includegraphics[width = 7.5 cm]{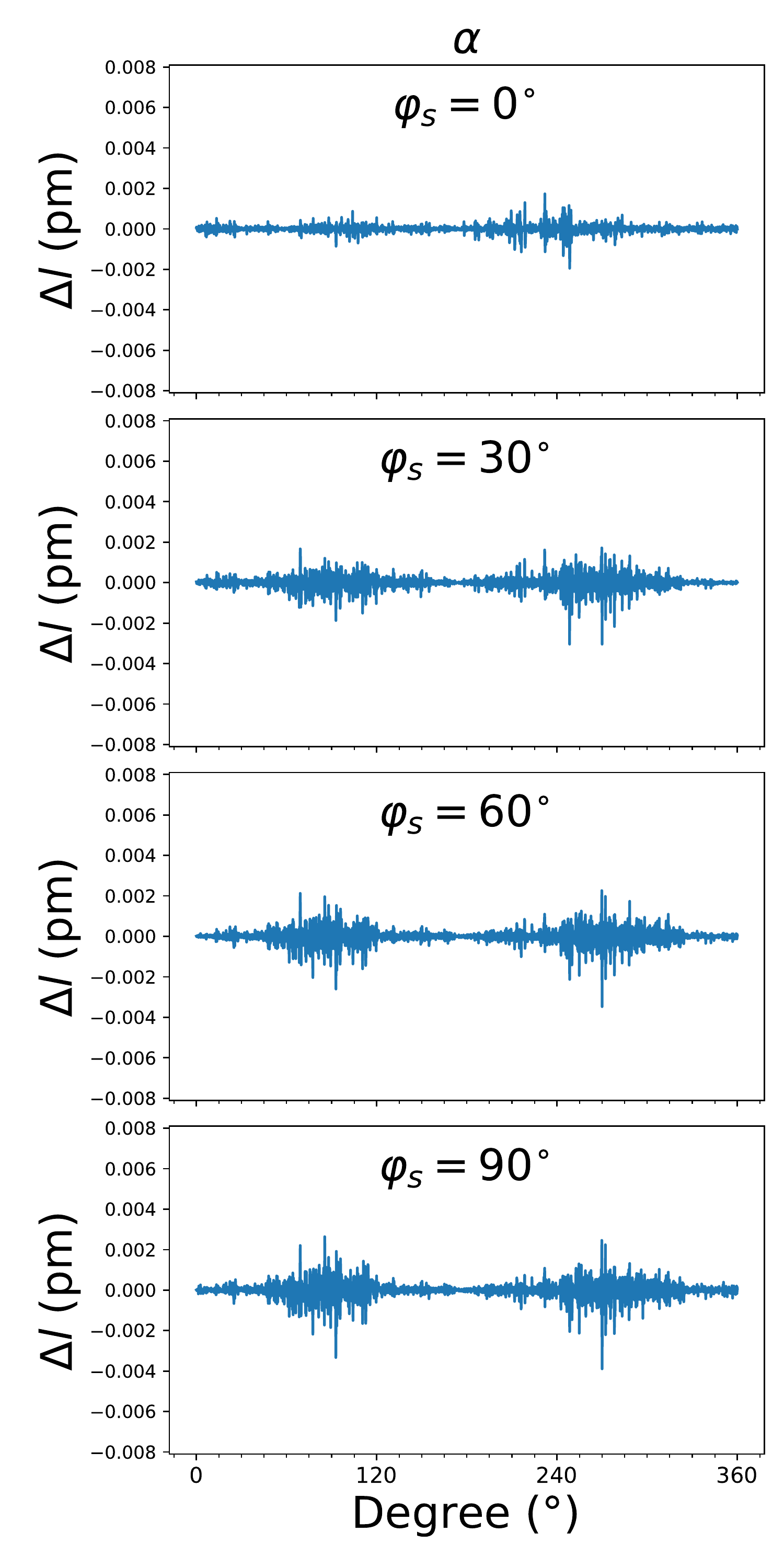}
		\end{minipage}
		\begin{minipage}[b]{0.419\textwidth}
			\includegraphics[width = 7.5 cm]{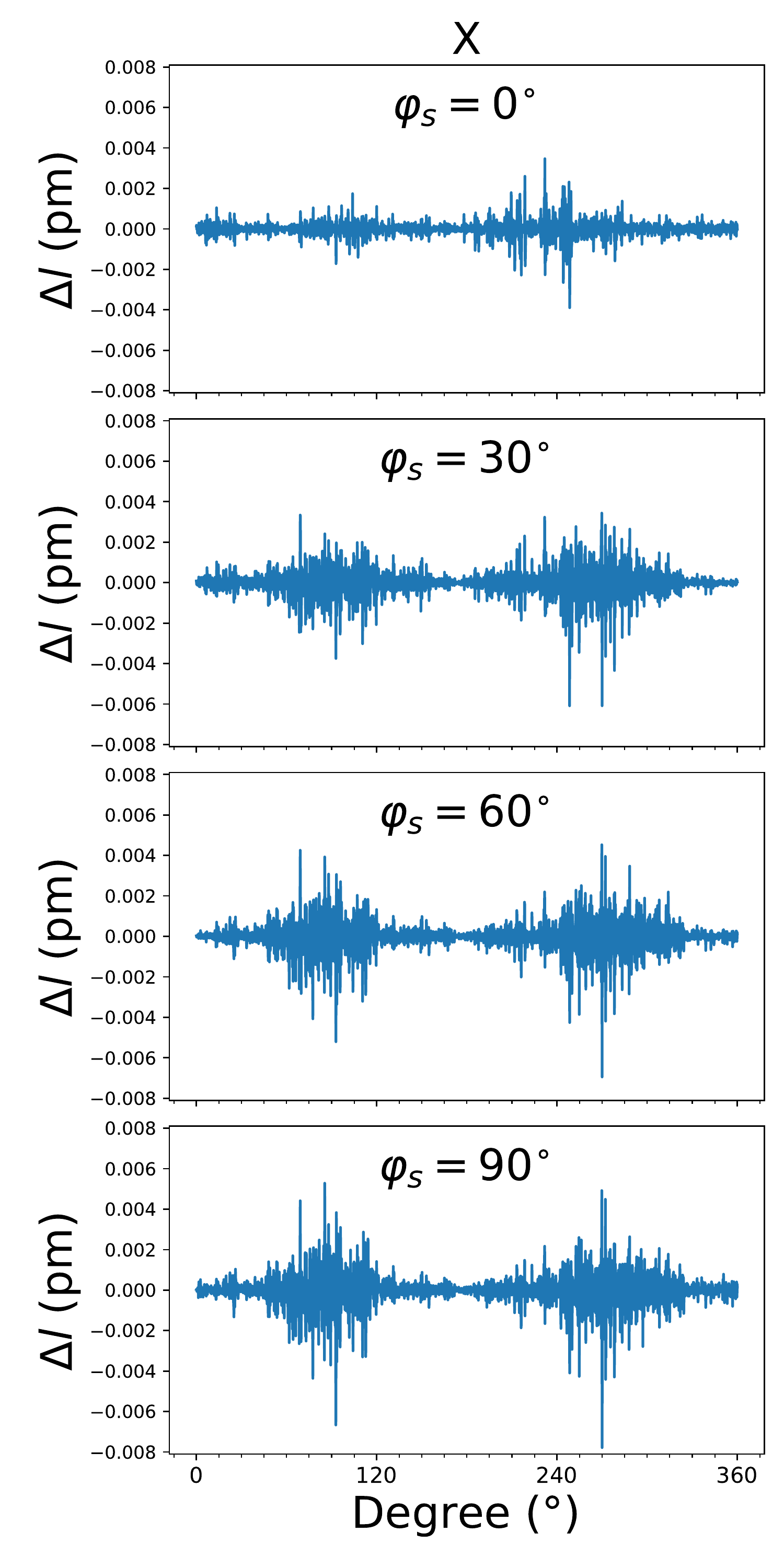}
		\end{minipage}
	\end{minipage}
	\caption{The OPD noises distributions for the $\alpha$ (left column) and $X$ (right column) combinations 
	on the orbit planes with $\varphi_s = 0^{\circ}$, $30^{\circ}$, $60^{\circ}$, and $90^{\circ}$. }
	\label{figDistribTDI}
\end{figure}

From Equation (\ref{Eq18}), we can see that the OPD noise reduction for the $\alpha$ combination 
is a factor of two better than that for the $X$ combination. 
This is because the laser beam will pass through one of the arms twice for the $X$ combination, 
but only once for the $\alpha$ combination (see Figure~\ref{figTDIpath}). 
Combining Equations (\ref{Eq12}), (\ref{Eq17}), and (\ref{Eq18}), we can calculate the OPD noises for the $\alpha$ and $X$ combinations as shown in Figure \ref{figDistribTDI}. The maxima of $|\Delta l|$ for the $\alpha$ and $X$ combinations are about 0.004 and 0.008 pm, respectively, which are about two orders of magnitude smaller than that for the Michelson combination. 
This indicates that TDI can significantly suppress the common-mode OPD noise. 

%This is because that α combination contains all the three arms, but X combination only contains two arms, the electron number density of the missing arm for X combination is quite different from the other two, which will lead to the OPD noise cannot be well eliminated by X combination.

\begin{figure}[ht]
%	\figurenum{8}
	\centering
	\begin{minipage}[b]{\textwidth}
		\centering
		\begin{minipage}[b]{0.419\textwidth}
			\includegraphics[width = 7.5 cm]{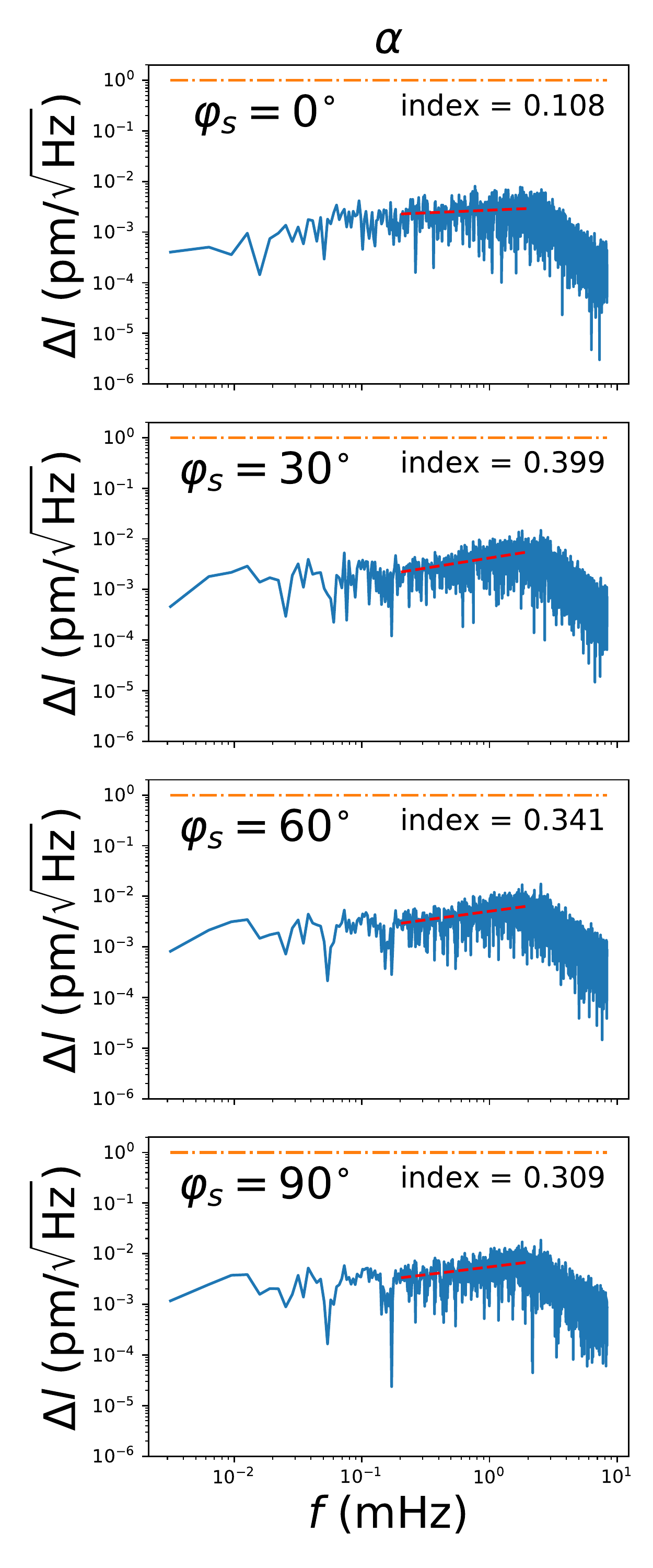}%width=1\textwidth
		\end{minipage}
		\begin{minipage}[b]{0.419\textwidth}
			\includegraphics[width = 7.5 cm]{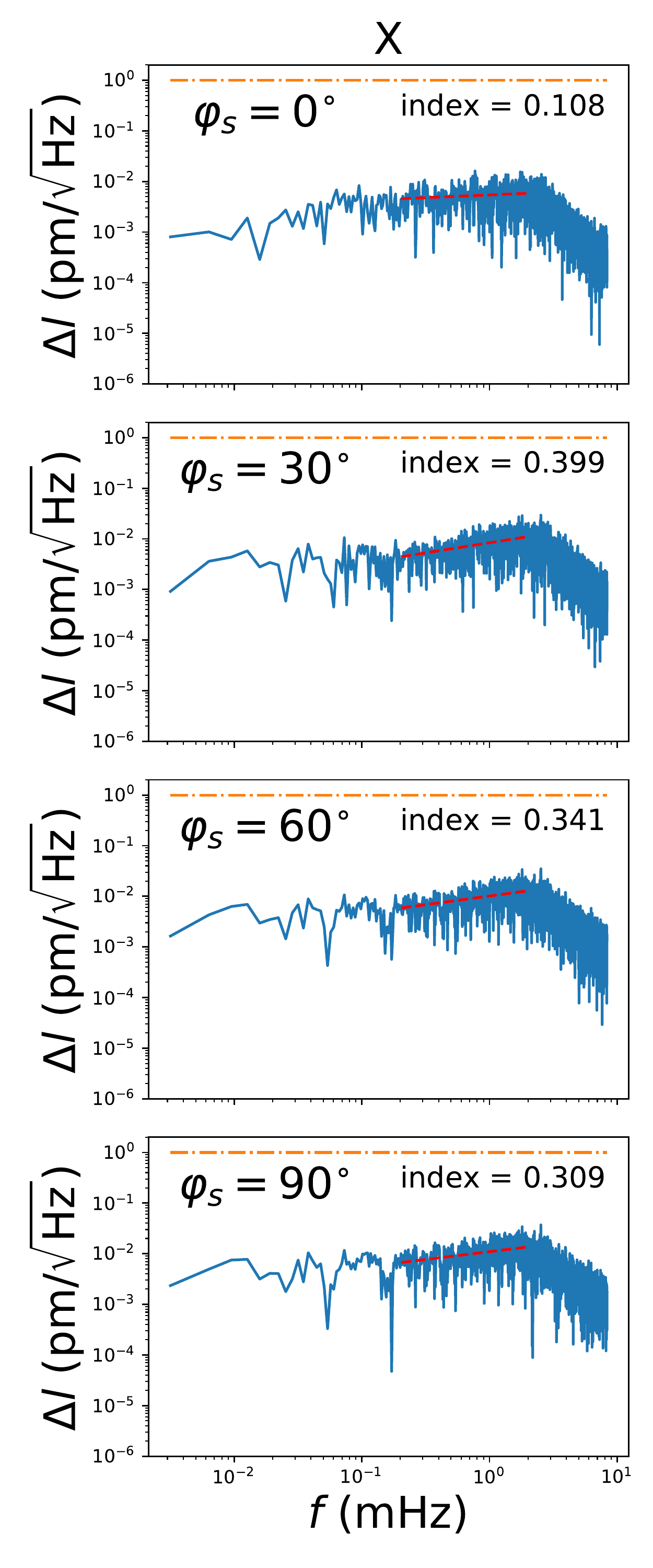}%width=1\textwidth
		\end{minipage}
	\end{minipage}
	
	\caption{The ASDs of the OPD noises for the $\alpha$ (left column) and X (right column) combinations on the orbit planes with $\varphi_s = 0^{\circ}$, $30^{\circ}$, $60^{\circ}$, and $90^{\circ}$. The orange dashed line is the proposed displacement measurement accuracy of TQ (1 pm/$\rm{\sqrt{Hz}}$) \citep{Luo2016}. Red dashed lines are the best-fit spectra of the ASDs. }
	\label{figASDTDI}
\end{figure}

Figure~\ref{figASDTDI} shows the ASDs of $\Delta l$ for the $\alpha$ and $X$ combinations. 
Similar to the single link and Michelson combination, the spectra of the ASDs become steeper when $f \gtrsim 1/800$ Hz. 
The best-fit spectral indices for the $\alpha$ ($X$) combination 
%are 0.075 (0.043), 0.364 (0.330), 0.307 (0.275), and 0.275 (0.242) 
are 0.108, 0.399, 0.341, and 0.309 
for $\varphi_s = 0^{\circ},~30^{\circ},~60^{\circ}$, and $90^{\circ}$; 
The corresponding spectral amplitudes at 1~mHz read 0.003 (0.005), 0.004 (0.008), 0.005 (0.010), 
and 0.005 (0.010) pm/$\sqrt{\rm Hz}$ at 1 mHz, respectively.

\section{Discussions} \label{discussions}

\subsection{The impact of OPD noise on the sensitivity} 

%Comparing $\Delta l$ for Michelson combination and the TDI combinations, we find that TDI combinations ($\alpha$ and $X$) can reduce the OPD noise significantly, especially $\alpha$ combination. 
%The ASDs of $\Delta l$ for Michelson combination show that it is a colored noise.
%However, TDI combinations make the OPD noises to be roughly like a white noise

\begin{figure}[ht!]
%	\figurenum{9}
	\plotone{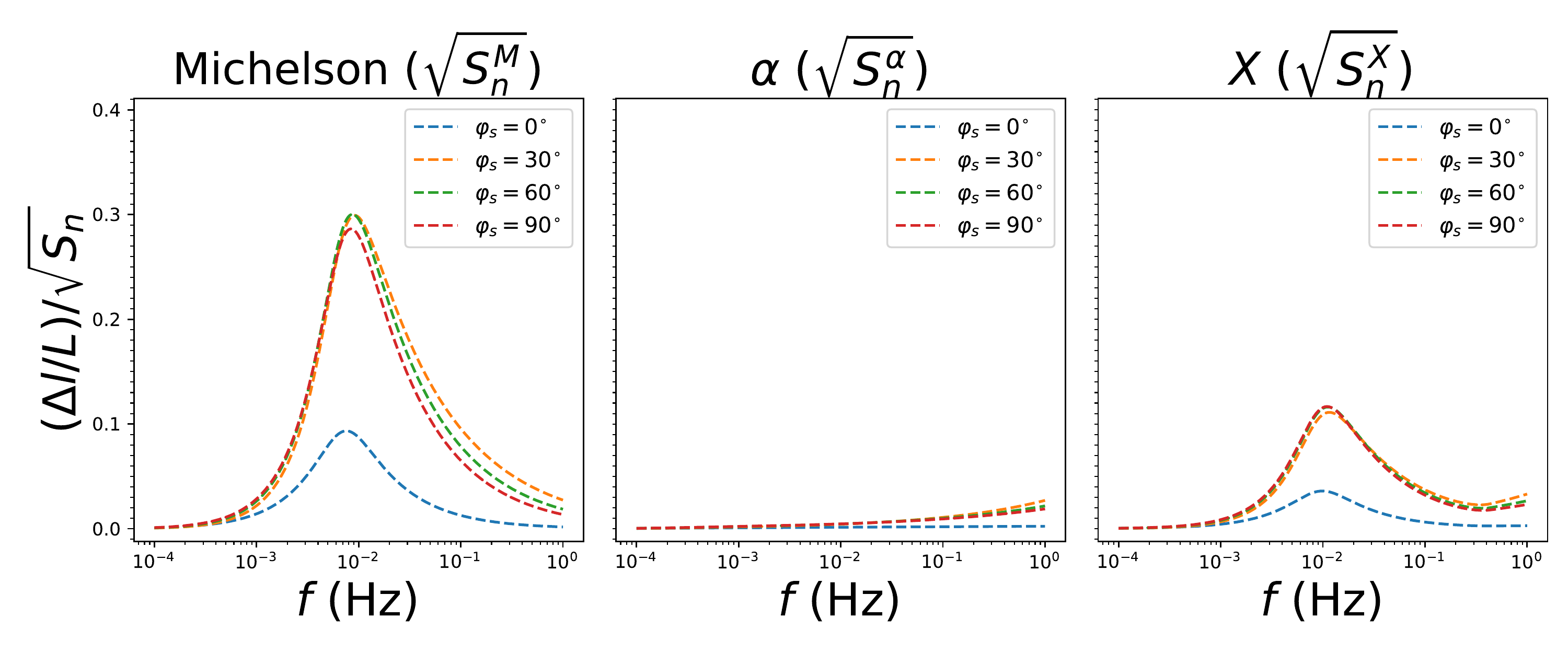}
	\caption{The ratios between the equivalent strain of the OPD noises and  that of TQ $\sqrt{S_n}$, i.e., $(\Delta l/L)/\sqrt{S_n^M}$ for the Michelson combination (left panel), $(\Delta l/L)/\sqrt{S_n^{\alpha}}$ for the TDI $\alpha$ combination (middle panel), and $(\Delta l/L)/\sqrt{S_n^X}$ for the TDI $X$ combination (right panel) on the orbit planes with $\varphi_s = 0^{\circ}$, $30^{\circ}$, $60^{\circ}$, and $90^{\circ}$. 
		}
	\label{figRatio}
\end{figure}

The equivalent strain noise ASD ($\sqrt{S_n}$) for the Michelson combination (denoted as $\sqrt{S_n^M}$) is as follows \citep{Hu2018},
\begin{equation}
\label{Eq19}
\begin{aligned}
S_n^M = S_n^x + S_n^a(1+\frac{10^{-4}~\mathrm{Hz}}{f} ) \,,
\end{aligned}
\end{equation}
where $\sqrt{S_n^x}$ is the equivalent strain noise ASD of the displacement measurement, $\sqrt{S_n^a}$ is the equivalent strain noise due to residual acceleration. 
And the equivalent strain noise ASD for the $\alpha$ ($\sqrt{S_n^{\alpha}}$) and $X$ ($\sqrt{S_n^{X}}$) combinations are as follows \citep{Armstrong1999},
\begin{equation}
\label{Eq20}
\begin{aligned}
S_n^{\alpha} = [4\mathrm{sin}^2(3 \pi f L/c) + 24\mathrm{sin}^2(\pi f L/c)]S_n^a + 6S_n^x \,,
\end{aligned}
\end{equation}
\begin{equation}
\label{Eq21}
\begin{aligned}
S_n^{X} = [4\mathrm{sin}^2(4 \pi f L/c) + 32\mathrm{sin}^2(2 \pi f L/c)]S_n^a + 16\mathrm{sin}^2(2\pi f L/c) S_n^x \,,
\end{aligned}
\end{equation}
In order to compare the OPD noises with TQ's equivalent strain noise ASDs for the Michelson ($\sqrt{S_n^M}$), $\alpha$ ($\sqrt{S_n^{\alpha}}$), and $X$ ($\sqrt{S_n^X}$) combinations, 
we calculate the ASDs of the equivalent strains of the OPD noises as $\Delta l/L$ for the corresponding data combinations. Using the best-fit spectra of the OPD noises for the Michelson, $\alpha$, and $X$ combinations, 
we calculate the ratio between $(\Delta l/L)$ and 
$\sqrt{S_n^M}$, $\sqrt{S_n^{\alpha}}$, $\sqrt{S_n^X}$, and the results are shown in Figure~\ref{figRatio}.

For the Michelson combination, the maximum of ($\Delta l/L)/\sqrt{S_n^M}$ is about 0.29 at $\approx 10$~mHz. 
($\Delta l/L)/\sqrt{S_n^M}$ increases with increasing frequency $f$ when $f \lesssim 10$ mHz. This is due to the fact that $\sqrt{S_n}$ is dominated, in this frequency range, by the acceleration noise with %the spectral index of -1 
$\sqrt{S_n^a}\propto f^{-2}$, the spectral index of which is less than the ones of $\Delta l$ for the Michelson combination. 
($\Delta l/L)/\sqrt{S_n^M}$ decreases with $f$ when $f \gtrsim 10$ mHz, 
and ($\Delta l/L)/\sqrt{S_n^M}$ decreases to about 0.04 at the transfer frequency $f_{\ast} \approx 0.3$~Hz \citep{Hu2018}. This is because $\sqrt{S_n^M}$ is dominated by the position noise which is approximately white in this frequency range. 

For the TDI $\alpha$ combinations, ($\Delta l/L)/\sqrt{S_n^{\alpha}}$ at $f \approx$ 10 mHz is about 0.005, 
which is about 1/60 of that for the Michelson combination.
However, as the spectral index of $\Delta l$ is slightly larger than 0 for $\alpha$ combinations, ($\Delta l/L)/\sqrt{S_n^{\alpha}}$ increase smoothly in the frequency range of $10^{-2} < f < 1$ Hz, and the maxima of ($\Delta l/L)/\sqrt{S_n^{\alpha}}$ from different $\varphi_{s}$ are about 0.015 at the transfer frequency $f_{\ast}$, which is much lower than that for the Michelson combination.
For the TDI $X$ combinations, 
%($\Delta l/L)/\sqrt{S_n^X}$ is much smaller than that for Michelson combination ($\Delta l/L)/\sqrt{S_n^M}$, 
the maximum of ($\Delta l/L)/\sqrt{S_n^X}$ is about 0.12 at $f \approx 10$~mHz, and ($\Delta l/L)/\sqrt{S_n^X} \approx 0.02$ at the transfer frequency $f_{\ast}$, which is also much lower than that for the Michelson combination, but higher than that for the $\alpha$ combination. ($\Delta l/L)/\sqrt{S_n^X}$ decreases with $f$ when $f \gtrsim 10$ mHz, this is because that the spectral index of $\sqrt{S_n^X}$ is larger than that of the $\Delta l$ for $X$ combination in the frequency range $f \gtrsim 10$ mHz.
The results suggest that the TDI combinations, especially $\alpha$ type,
can efficiently suppress the common-mode OPD noise produced during the laser propagation in space plasma for the most sensitive frequency range of TQ. % ($f\lesssim 0.1$ Hz). 
However, due to the limited high frequency reach of our simulation, the results in Figure~\ref{figRatio}, which are obtained by extrapolating the red dashed lines in Figure~\ref{figASDTDI}, may overestimate or underestimate the OPD noises when $f \gtrsim 1/800$~Hz. Further investigation with higher spatial and temporal resolution is needed. 

Since the spectral indices of the ASDs of the OPD noises for a single link and a Michelson combination are about -2/3, the OPD noise will be more significant in lower frequencies. And considering that LISA is more sensitive than TQ in lower frequencies ($f \lesssim$ 10 mHz) \citep{PhysRevD.95.103012,Feng2019}, the ratios between $(\Delta l/L)$ and $\sqrt{S_n}$ of LISA in the lower frequencies for the single link will be larger than that of TQ.
Howerver, in practice, for LISA, TQ and other space-borne GWs detectors, 
%the laser ranging combinations are TDI combinations rather than Michelson or single link. Fortunately, the TDI combinations can suppress the OPD noise significantly in the most sensitive frequency range of TQ.
one will use TDI data combinations which can suppress the otherwise dominating laser frequency noise, rather than single link data, in GW data analysis. Based on our results, TDI can partially cancel the common-mode OPD noise. 
%and make the spectral indices of the ASDs of the OPD noises to be about 1/3.
	
%In addition, as shown in Fig. \ref{figDistribS}, \ref{figDistribM} and \ref{figDistribTDI}, the OPD noise is obviously non-stationary for the single link, Michelson, and TDI combinations. Based on our results, the OPD noise rooted from space plasma, especially for these TDI combinations, are not a dominating source. Therefore, it may not be able to change the overall statistical properties of the total noise. 
%The non-stationary noises can be addressed by using a B-spline method to modelling the noise PSD over time \citep{Mohanty2000,Edwards2020}.
In addition, Figures 3, 6 and 8 show respectively that the OPD noises rooted from space plasma are strongly non-stationary for the single link, Michelson, and TDI combinations. However, as shown in Figure 10, the OPD noises, especially for the TDI combinations, are not dominating in the total noise budget. Therefore, it may not significantly change the anticipated overall statistical properties of the total noise. On the other hand, when the non-stationarity in the total noise is strong, identifying and addressing the non-stationary noise will become crucial in the GW data analysis \citep{Mohanty2000,Edwards2020}. The non-stationarity in noise deserves careful consideration in the development of GW detectors. 

\subsection{Space weather}

The solar wind dynamic pressure $P_{\rm dyn}$ is the most important parameter that determines the Earth magnetosphere's geometric structures and distributions of electron number density \citep{Lv2015,Wang2016}. The more strongly the magnetosphere is compressed by the solar wind, and the larger the number density will be \citep{Lv2015}, which leads to the larger amplitude of the OPD noise. 
From the OMNI data of the solar wind \citep{King2005}, we obtain $P_{\rm dyn}$ with the value of $2.0 \pm 1.2~{\rm nPa}$ during a total solar cycle from 1997 to 2019. In this work, the input $P_{\rm dyn} = 2.1 \pm 0.7~{\rm nPa}$, which is approximated to the mean value of $P_{\rm dyn}$ during the total solar cycle. 
Consider that both the period of $P_{\rm dyn}$ and solar magnetic cycle are about 22 years and TQ is proposed to launch at early 2030s, $P_{\rm dyn}$ that is obtained 22 years before the real operation is a good approximation. We get $P_{\rm dyn}$ data from 2008 to 2012 (about 22 years before the lanuch) on OMNI website, and find that the time when $P_{\rm dyn}$ is larger than 3 nPa and 5 nPa accounts for only about 5\% and 1\% of the total period. 
Besides, the Earth magnetoshpere encounters the solar eruptions, e.g. interplanetary shocks and coronal mass ejections, $P_{\rm dyn}$ can reach or even exceed 5 nPa in these cases.
The laser propagation noises in the cases of solar eruptions are expected in future works.
Recently, the OPD noise of LISA was estimated based on the in-situ observations from the Wind spacecraft \citep{Smetana2020}. 
%MHD simulations are popular in the studies of heliophysics, it can avoid the shortcoming of in-situ observations without spatial resolution, and it can get the global structure around the laser beams of GW detectors. 
%In contrast to the in-situ observations, the MHD simulation can provide the global structure of the space environment with spatial resolution around the GW orbit and along the laser beams. 
%	Thus, MHD simulations can be considered as a more appropriate approach in the future studies of the laser propagation noise for other GW detectors, e.g. LISA, Taiji, BBO, DECIGO. 
In contrast to the in-situ observations, the MHD simulation with the input of real time solar wind data can provide the global structure of the space environment with spatial resolution along the laser beams, and it can be applied to the future investigations of laser propagation noise for the other space-borne GW detectors. 
%Thus, MHD simulations with possible in-situ data as the input can serve as an important complimentary approach for future studies of the laser propagation noise for other GW detectors, e.g. LISA, Taiji, BBO, DECIGO. 
Furthermore, the SWMF model that used in this work is an MHD model, only the physical processes at global and the MHD scales can be revealed. 
On the other hand, the plasma-scale physical processes, such as plasma waves and turbulences, cannot be revealed by the SWMF model. For the impact of plasma scale physical process on the OPD noise, the hybrid or particle-in-cell (PIC) simulations are needed.

Besides obtaining the TEC from MHD simulations, it is also possible to derive the TEC from real-time observations.
In order to obtain the TEC along the laser propagation path, we can transmit signals with two frequencies to reduce $\Delta l$, the dual-frequency scheme has been used in the Compass system \citep{Yang2011}, the Gravity Recovery and Climate Experiment (GRACE) and GRACE Follow-on (GRACE-FO) \citep{Tapley2004,Landerer2020}.
For dual-frequency scheme, there are two general methods, one is differential group delay, the other is differential carrier phase. Differential group time delay measures the time delay difference of two EM waves with different frequencies, and differential carrier phase measures the phase difference of two EM waves with different frequencies. The accuracy of differential group delay method is lower than that of differential carrier phase, but it can measure the absolute value of TEC. 
For the next-generation space-borne GW detectors, e.g., DECIGO \citep{DECIGO2011}, the proposed strain sensitivity is about five orders of magnitude lower than that of TQ, and the difference of the electron density between the solar wind at 1 AU and that around the TQ orbit with a geocentric altitude of $10^5$ km is generally no more than one order of magnitude. Thus, the laser propagation noise will become a dominating environmental noise for DECIGO, and the laser ranging scheme with the dual-frequency laser will become a necessity. 
%and transmitting two frequency signals can be considered as an alternative scheme to eliminate the OPD noise.
%The two frequency scheme can be used as a scheme for the next generation of space-based GW detectors to solve the OPD noise caused by space plasma.

\section{Conclusions} \label{conclusions}

Dispersion can cause OPD noise when the laser beams propagate in space plasma. 
In this work, the Appleton-Hartree equation, the orbits of TQ satellites and the global magnetosphere simulation based on the SWMF 
are used to analyze the OPD noise $\Delta l$ for TQ at four typical relative positions of the detector's planes 
with $\varphi_s = 0^{\circ}, 30^{\circ}, 60^{\circ}$, and $90^{\circ}$. 

The maxima of $|\Delta l|$ for the single link and the Michelson combination are about 1 and 3 pm, respectively. The maxima of $|\Delta l|$ can be reduced to about 0.004 and 0.008 pm for the TDI combination $\alpha$ and $X$, respectively. 
%The ASDs of OPD noise for Michelson combination suggest that it is a colored noise, and TDI combinations make the OPD noise to be roughly like a white noise.
Furthermore, we calculate the ratio between the equivalent strain of the OPD noise and the one proposed for TQ, i.e., $(\Delta l/L)/\sqrt{S_n^M}$, $(\Delta l/L)/\sqrt{S_n^{\alpha}}$, $(\Delta l/L)/\sqrt{S_n^X}$ for the Michelson, $\alpha$, $X$ combinations, respectively. We find that in the most sensitive frequency range of TQ, the TDI combinations can suppress the OPD noise significantly. %but in the high frequencies with $f \gtrsim 0.1$ Hz, the OPD noise for the Michelson and TDI combinations are on the same order of magnitude.
For the next-generation space-borne GW detectors, transmitting EM waves with two frequencies can be considered 
as a necessity to significantly reduce the OPD noise.

\acknowledgments

The simulations used in this work are provided by the Community Coordinated Modeling Center (CCMC) at Goddard Space Flight Center through their public Runs on the Request system. This work is carried out using the SWMF and BATSRUS tools developed at the University of Michigan's Center for Space Environment Modeling.
S.W. is supported by National Key R \& D Program of China (Grant 2020YFC2201201), the National Natural Science Foundation of China (NSFC) under grant 11803008. 
Y.W. is supported by NSFC under grants 11973024 and 11690021, and Guangdong Major Project of Basic and Applied Basic Research (Grant No. 2019B030302001). 
L.L.F. is supported by NSFC under grant 42004156.
S.T. was supported by NASA grant 80NSSC17K0453.
Z.Z.B. is supported by NSFC under grant 11727814 and 11975105.
C.P.F. is supported by NSFC under grant 11961131002.

\bibliography{sample63}{}
\bibliographystyle{aasjournal}

\end{CJK*}
\end{document}